\documentclass[conference]{IEEEtran}
\IEEEoverridecommandlockouts

\usepackage{graphicx}
\usepackage{amsmath,amssymb,amsthm}
\usepackage{algorithm,algorithmic} 
\usepackage{multirow}
\usepackage{tikz}
\usepackage{tikz-3dplot}
\usepackage[caption=false,font=footnotesize]{subfig}
\usetikzlibrary{arrows.meta,positioning,calc,fit}
\definecolor{HBlue}{RGB}{55,126,184}   
\definecolor{VRed}{RGB}{228,26,28}     
\definecolor{Gray70}{RGB}{70,70,70}    
\colorlet{PurpleMix}{HBlue!70!VRed}    
\usepackage{cite}
\usepackage{url}
\usepackage{balance} 
\usepackage{xcolor}
\usepackage{dsfont}

\usepackage{multirow}
\usepackage[table]{xcolor}
\usepackage{booktabs,tabularx,makecell,array}
\usepackage[hidelinks]{hyperref}

\newcommand{\Tau}{\mathrm{T}}
\newcommand{\Tad}{$\theta$-adaptive}
\newcommand{\Tag}{$\theta$-agnostic}
\title{ A Three-Dimensional Path Loss Model for THz Band Aerial Communications}
\author{
Sina Jorjani$^{1}$, Caglar Tunc$^{1}$, Ozgur Gurbuz$^{1}$, Akhtar Saeed$^{2}$ \\
$^{1}$Faculty of Engineering and Natural Sciences, Sabancı University, İstanbul, Türkiye \\
$^{2}$Department of Electrical-Electronics Engineering, Kadir Has University, İstanbul, Türkiye \\
\{sina.jorjani, caglar.tunc, ogurbuz\}@sabanciuniv.edu, akhtar.saeed@khas.edu.tr \\
\thanks{This work was supported in part by the Scientific and Technological Research Council of Türkiye (TÜBİTAK) under Grant No. 224N014 and by Horizon Europe under MSCA-2024-PF-01 Project No. 10120773.}
}

\begin{document}
\maketitle
\begin{abstract}
Accurate characterization of Terahertz (THz) band path loss is critical for reliable high-frequency communication, especially in aerial networks where transceivers may operate at different altitudes.
Existing THz-band path loss models for aerial networks focus on horizontal or vertical transceiver deployments, and fall short at modeling the random 3D geometry of transceiver locations. 
To address this limitation, we propose a new analytical THz path loss model that incorporates arbitrary 3D geometry of transceiver locations and frequency-selective absorption, obtained through a two-dimensional regression. We validate our proposed model with the propagation data collected via the Atmospheric Model (am) tool for multiple aerial link types, including drone-to-drone (Dr2Dr), medium-altitude aerial communication (MAAC), high-altitude unmanned aerial vehicles~(UAV)-to-UAV (U2U) links over varying transceiver separation and sub-THz to low-THz spectrum, i.e., 0.1--1~THz. The proposed framework provides a unified and accurate model for analyzing and designing future high-frequency aerial communication systems.
\end{abstract}


\vspace{-2mm}
\section{Introduction} 
With the emergence of ultra–high data rate applications such as digital twins, virtual and augmented reality, 3D streaming, and the rapid growth in the number of connected devices, the migration toward high-frequency bands, such as Terahertz (THz) and sub-THz, has become inevitable. In this regard, the THz band (0.1--10~THz) has been identified as a strong candidate for 6G communication systems, due to its ability to provide extremely high data throughput and spatial resolution by offering a broader spectrum than the mmWave band and leveraging both electromagnetic and optical waves~\cite{6GZhang}.

Despite the promising features of the THz band communication in 6G networks, radio waves experience substantial loss at the THz and sub-THz frequency ranges, both as a result of spreading loss and atmospheric absorption loss, which is mainly due to the water vapor and oxygen molecules in the air. The absorption characteristics in these frequencies can vary significantly with frequency, atmospheric conditions, and environmental factors, such as fog and rain, which can severely limit the effective communication distance~\cite{liao2023attenuation,taleb2023propagation,saeed2024terahertz}.

The limitations introduced by absorption loss are typically the most severe in terrestrial communications and ground-based systems, as the water vapor density is highest at lower altitudes~\cite{SAEED_Phycom}. Therefore, the development and deployment of communication links for non-terrestrial network (NTN) layers, such as satellites, high-altitude platforms (HAPS), and unmanned aerial vehicles (UAVs), is crucial, as the communication over the THz band becomes significantly more reliable when operating above the denser layers of the atmosphere \cite{TMao2022}.

Recent studies have examined the distinct propagation characteristics of terrestrial and aerial THz links, yet research in this domain remains limited.  
Several works have modeled path loss and absorption for THz and sub-THz frequencies to simplify the frequency-selective nature of atmospheric attenuation. In~\cite{kokkoniemi2018simplified}, a polynomial-fitting model for molecular absorption loss between 275--400~GHz was proposed, accurately reproducing line-by-line results within 1~km ranges. Expanding on this,~\cite{danobrega2023channel} introduced a broadband channel loss model spanning 100--600~GHz that integrates discrete molecular lines and the \textit{water vapor continuum}, derived from HITRAN spectroscopic data. The model closely matches high-resolution computations while offering a closed-form analytical framework suitable for 6G THz systems evaluations.

For aerial and non-terrestrial environments,~\cite{AnalyticalModel_paper} proposed a regression-based analytical path loss model for low-altitude Drone-to-Drone (Dr2Dr) links across two transmission windows in the 790--940~GHz range, assuming horizontally aligned drones. The model compactly relates transmittance to altitude, distance, and frequency. In~\cite{khan2025analysis}, the authors showed that high-altitude airborne links experience notably lower absorption than ground-based paths, enabling airplane-to-satellite 
links to sustain reliable connectivity up to 1~THz, while ground-to-satellite 
links experience strong low-altitude absorption when operating above a few hundred GHz. Similarly, Doborshchuk \textit{et al.}~\cite{doborshchuk2022propagation} developed an analytical 0.3--3~THz ground-to-aircraft model incorporating altitude-dependent and weather-induced attenuation, showing that adverse conditions such as heavy rain can add up to 50~dB of excess loss.

Despite such advances, analytical and empirical models remain scenario-specific and geometrically constrained, focusing on vertical or horizontal settings, such as \cite{AnalyticalModel_paper}.
On the other hand, many studies are designed to account for atmospheric negative effects like clouds, rain, and fog as in~\cite{khan2025analysis,danobrega2023channel}. 
While these studies are undoubtedly valuable, they fail to provide a comprehensive view of the path loss behavior in a more general sense. In this context, this paper aims to fill this gap by generalizing the aerial path loss formulation to arbitrary 3D transceiver geometries over ten representative THz sub-bands in the 0.1--1~THz range, while looking at a wider range of operational altitudes. 
The main contributions of the paper are summarized as follows:
\begin{itemize}
\item The proposed model encompasses multiple aerial communication scenarios characterized by a random 3D geometry of transceiver locations, including Dr2Dr, medium-altitude aerial communication (MAAC), and high-altitude unmanned aerial (U2U) links.
\item The proposed path loss model is evaluated over a wide range of low-THz and sub-THz frequency bands (0.1--1~THz), with detailed analysis across ten representative sub-bands to capture frequency-dependent propagation characteristics.
\item The model is generalized in terms of geometry, allowing transceivers to be located at arbitrary 3D positions with respect to each other, rather than being limited to purely horizontal or vertical link configurations.
\end{itemize}

The rest of the paper is organized as follows.
In Section~\ref{section_sysModel}, the system model and the proposed path loss estimation model are introduced. Section~\ref{section_ImpApp} explains Atmospheric Model (\textit{am}) data acquisition and develops two different implementation approaches for path loss estimation. Next, the proposed methods are simulated and the results are discussed in section~\ref{section_Results}. Finally, Section~\ref{section_Conclusion} concludes the paper.

\section{System Model} \label{section_sysModel}
We consider a communication scenario with two transceivers, \textit{e.g.}, drones $\mathrm{N}_1$ and $\mathrm{N}_2$, located at coordinates $(x_1, y_1, z_1)$ and $(x_2, y_2, z_2)$, respectively, in a Cartesian coordinate system, where the $z = 0$ meters plane indicates sea level. The vertical distance between the transceivers is defined as $d_v = |z_2 - z_1|$, and the horizontal distance is calculated as
$d_h = \sqrt{(x_2 - x_1)^2 + (y_2 - y_1)^2}$, both in meters.
Therefore, the total distance between the transmitter and the receiver in meters can be expressed as
$d = \sqrt{d_h^2 + d_v^2}$.

\begin{figure}[t]
\centering
\tdplotsetmaincoords{70}{110} 
\begin{tikzpicture}[tdplot_main_coords, scale=1]

\coordinate (Dr1) at (0.5, 0.5, 1);         
\coordinate (Dr2) at (-2, 2, 2);            
\coordinate (Dr2proj) at (-2, 2, 1);        

\draw[->] (0,0,0) -- (4,0,0) node[anchor=north east]{$x$};
\draw[->] (0,0,0) -- (0,3,0) node[anchor=north west]{$y$};
\draw[->] (0,0,0) -- (0,0,2) node[anchor=south]{$z$};

\filldraw[tdplot_main_coords, fill=red!18, opacity=0.3, draw=none] 
  (-1,-1,1) -- (3,-1,1) -- (3,2,1) -- (-1,2,1) -- cycle;

\filldraw[tdplot_main_coords, fill=blue!18, opacity=0.3, draw=none] 
  (-1,-1,0) -- (3,-1,0) -- (3,2,0) -- (-1,2,0) -- cycle;

\node[tdplot_main_coords, anchor=center] at (1.5,1.5,0) {\footnotesize Sea level ($z=0$)};

\begin{scope}[shift={(Dr1)}]
  \fill[black] (0,0,0) circle (0.07); 
  \draw[thick] (-0.21,0,0) -- (0.21,0,0); 
  \draw[thick] (0,-0.21,0) -- (0,0.21,0);
  \fill[gray] (-0.21,0,0) circle (0.055); 
  \fill[gray] (0.21,0,0) circle (0.055);
  \fill[gray] (0,0.21,0) circle (0.055);
  \fill[gray] (0,-0.21,0) circle (0.055);
  \node[above] at (0,0,0.07) {$\mathrm{N}_1$}; 
\end{scope}

\begin{scope}[shift={(Dr2)}]
  \fill[black] (0,0,0) circle (0.07);
  \draw[thick] (-0.21,0,0) -- (0.21,0,0);
  \draw[thick] (0,-0.21,0) -- (0,0.21,0);
  \fill[gray] (-0.21,0,0) circle (0.055);
  \fill[gray] (0.21,0,0) circle (0.055);
  \fill[gray] (0,0.21,0) circle (0.055);
  \fill[gray] (0,-0.21,0) circle (0.055);
  \node[above right] at (0,0,0.07) {$\mathrm{N}_2$};
\end{scope}

\draw[dashed, thick] (Dr1) -- (Dr2) node[midway, above] {$d$};
\draw[<->] (Dr1) -- (Dr2proj) node[midway, below right] {$d_h$};
\draw[<->] (Dr2) -- (Dr2proj) node[midway, right] {$d_v$};
\draw[<->] (Dr1) -- (0.5,0.5,0) node[midway, right] {$z_1 = l$};

\def\phivec{-30.96}
\def\thetavec{108.93}
\begin{scope}
  \tdplotsetrotatedcoordsorigin{(Dr2)}
  \tdplotsetthetaplanecoords{\phivec}
  \tdplotdrawarc[tdplot_rotated_coords, thick, ->]
    {(0,0,0)}{0.3}{\thetavec}{180}
    {anchor=north, xshift=-2pt, yshift=2pt}{$\theta$};
\end{scope}
\end{tikzpicture}
\caption{Dr2Dr communication scenario in 3D, showing distances $d$, $d_h$, and $d_v$, zenith angle $\theta$, and the sea level at $z = 0$. 
}
\label{fig:drone_scenario}
\vspace{-4mm}
\end{figure}
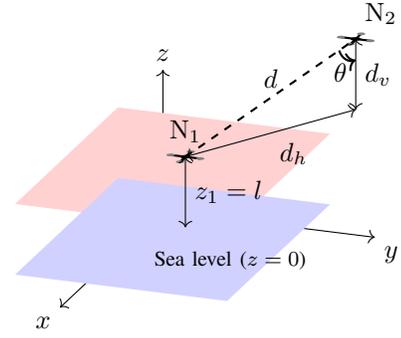

Let $\theta \in [0^\circ,90^\circ]$ denote the zenith angle between $\mathrm{N_1}$ and $\mathrm{N_2}$. $\theta = 0^\circ$ corresponds to the vertical case (\textit{i.e.}, $N_2$ is located directly above $N_1$), while $\theta = 90^\circ$ corresponds to the horizontal case. The horizontal and vertical distances in meters are then given by $d_h = d \sin{\theta}$, and $d_v = d \cos{\theta}$, respectively,
as illustrated in Fig.~\ref{fig:drone_scenario}, for the Dr2Dr scenario.

The total path loss $L_{\mathrm{total}}$ consists of two main components: the Free-Space Path Loss (FSPL) $L_{\mathrm{fspl}}$ and the absorption loss $L_{\mathrm{abs}}$, which is mainly due to the concentration of water vapor:
\begin{equation} \label{eq_pathLoss_1}
   L_{\mathrm{total}}(z_1,z_2,d,f) = L_{\mathrm{fspl}}(d,f) + L_{\mathrm{abs}}(z_1,z_2,d,f),
\end{equation}
where $f$ denotes the carrier frequency, and $ L_{\mathrm{total}}$, $L_{\mathrm{fspl}}$, and $L_{\mathrm{abs}}$ are in $\mathrm{dB}$. Without loss of generality, let $l = z_1 \leq z_2$; we can uniquely identify $(d, z_1, z_2, f)$ with $(d_h, d_v, l, f)$ and vice versa. Thus, \eqref{eq_pathLoss_1} can be redefined as
\vspace{-0.5mm}
\begin{equation} 
\label{eq_pathLoss_2}
   L_{\mathrm{total}}(d_h, d_v, l, f) = L_{\mathrm{fspl}}(d, f) + L_{\mathrm{abs}}(l, d_h, d_v, f),
\end{equation}
where $L_{\mathrm{fspl}}(d, f) = 20\log_{10}\left(\frac{4\pi f d}{c}\right)$, with $c \approx 3\cdot10^8\,$ m/s denoting the speed of light in free space. The absorption loss is given by $L_{\mathrm{abs}}(d_h, d_v, l, f) = -10\log_{10}\tau(d_h, d_v, l, f)$, where $\tau(\cdot) \in (0,1)$ represents the transmittance, defined as the inverse of the absorption coefficient~\cite{jornet2011channel}.

In this work, we adopt a data-driven approach, inspired by~\cite{AnalyticalModel_paper}, to analytically model THz-band path loss as a function of horizontal and vertical distances, altitude, and frequency. The transmittance data used for model derivation and evaluation is obtained via the \textit{am} tool~\cite{am_Tool}. In our framework, transmittance is first modeled as an exponential function of horizontal and vertical distances at fixed frequency and altitude, with parameters obtained as follows:

\footnotesize
\begin{equation}\label{eq_cascadedReg_1}
\begin{aligned}
&\{\hat{b}_{1,h}(l,f), \hat{b}_{1,v}(l,f)\} =\\ &\mathop{\mathrm{arg\,min}}_{b_{1,h},\, b_{1,v}}
\sum_{i}\Bigl\|
\ln \!\bigl(\tau(l, \boldsymbol{d}_h[i], \boldsymbol{d}_v[i], f)\bigr)
- b_{1,h}\, \boldsymbol{d}_h[i]
- b_{1,v}\, \boldsymbol{d}_v[i]
\Bigr\|_2^2,
\end{aligned}
\end{equation}
\normalsize
where $\boldsymbol{d}_h$ and $\boldsymbol{d}_v$ are the vectors of horizontal distance and vertical distance samples, respectively. 
Subsequently, the approximated exponents $\hat{b}_{1,h}(l,f)$ and $\hat{b}_{1,v}(l,f)$ are modeled as exponential functions of altitude $l$ at a fixed frequency:
\footnotesize
\begin{equation}\label{eq_cascadedReg_2}
    \{\hat{a}_{2,\xi}(f), \hat{b}_{2,\xi}(f)\} =\mathop{\mathrm{arg\,min}}_{a_{2,\xi}, b_{2,\xi}}\sum_{j}\left\| \hat{b}_{1,\xi}(\boldsymbol{l}[j],f) - a_{2,\xi} e^{b_{2,\xi} \boldsymbol{l}[j]}\right\|_2^2,
\end{equation}
\normalsize
where $\xi \in \{h, v\}$, and $\boldsymbol{l}$ is the vector of altitude samples.

\begin{figure}[!t]
\centering
\begin{tikzpicture}[>=Latex, font=\scriptsize, node distance=5mm]
  \newcommand{\BlockW}{30mm} 
  \tikzset{
    block/.style={draw, rounded corners=2pt, inner sep=2pt, align=center, text width=\BlockW},
    blockH1/.style={block, fill=HBlue!25},
    blockH2/.style={block, fill=HBlue!25},
    blockH3/.style={block, fill=HBlue!25},
    blockV1/.style={block, fill=VRed!25},
    blockV2/.style={block, fill=VRed!25},
    blockV3/.style={block, fill=VRed!25},
    blockTau/.style={block, fill=PurpleMix!40},
    blockOut/.style={block, fill=PurpleMix!40, text width=48mm},
    hconn/.style={->, HBlue, line width=0.9pt},
    vconn/.style={->, VRed,  line width=0.9pt}
  }

  \node[blockTau] (tau) {$\tau(l,d_h,d_v,f)=\tau_h\,\tau_v$};

  \node[blockH1, below=8mm of tau, xshift=-19mm] (h1)
    {\textbf{Step 1 ($h$)}\\$\ \hat{\tau}_h(l,d_h,f)= \exp(\hat{b}_{1,h}(l,f)d_h)$};
  \node[blockH2, below=3.6mm of h1] (h2)
    {\textbf{Step 2 ($h$)}\\$\hat{b}_{1,h}(l,f)\!\approx\!\hat{a}_{2,h}(f)e^{\hat{b}_{2,h}l}$};
  \node[blockH3, below=3.6mm of h2] (h3)
    {\textbf{Step 3 ($h$)}\\$\hat{a}_{2,h}(f)\!\approx\!\sum_{p=0}^{P_h}\hat{\lambda}_{h}[p]f^{p}$};

  \node[blockV1, below=8mm of tau, xshift=+19mm] (v1)
    {\textbf{Step 1 ($v$)}\\$\hat{\tau}_v(l,d_v,f)= \exp(\hat{b}_{1,v}(l,f)d_v)$};
  \node[blockV2, below=3.6mm of v1] (v2)
    {\textbf{Step 2 ($v$)}\\$\hat{b}_{1,v}(l,f)\!\approx\!\hat{a}_{2,v}(f)e^{\hat{b}_{2,v}l}$};
  \node[blockV3, below=3.6mm of v2] (v3)
    {\textbf{Step 3 ($v$)}\\$\hat{a}_{2,v}(f)\!\approx\!\sum_{p=0}^{P_v}\hat{\lambda}_{v}[p]f^{p}$};

  \coordinate (splitL) at ($(tau.west)+(0,0)$);
  \coordinate (splitR) at ($(tau.east)+(0,0)$);

  \node[Gray70, fill=white, inner sep=1pt, anchor=east]
      at ($(splitL)+(-3mm,-5mm)$) {\scriptsize{ Horizontal ($h$) branch}};
  \node[Gray70, fill=white, inner sep=1pt, anchor=west]
      at ($(splitR)+(+2.7mm,-5mm)$) {\scriptsize{ Vertical ($v$) branch}};

  \draw[hconn] (splitL) to[out=210,in=90] (h1.north);
  \draw[vconn] (splitR) to[out=-30,in=90] (v1.north);

  \draw[hconn] (h1.south) -- (h2.north);
  \draw[hconn] (h2.south) -- (h3.north);
  \draw[vconn] (v1.south) -- (v2.north);
  \draw[vconn] (v2.south) -- (v3.north);

  \path let \p1=(h3), \p2=(v3) in coordinate (midbelow) at ($(\x1,0)!0.5!(\x2,0)$);
  \node[blockOut, below=50mm of midbelow] (out)
    {$\hat{\tau}(l,d_h,d_v,f)= \hat{\tau}_h\hat{\tau}_v = \exp\!\left(
  \hat{\Lambda}_h(f) e^{\hat{b}_{2,h} l}\, d_h \right.\left. +
  \hat{\Lambda}_v(f) e^{\hat{b}_{2,v} l}\, d_v
\right)$};
      \coordinate (join1) at ($(out.north)+(-2.5mm,0)$);
      \coordinate (join2) at ($(out.north)+(+2.5mm,0)$);
  \draw[hconn] (h3.south) to[out=-90,in=90,looseness=1] (join1);
  \draw[vconn] (v3.south) to[out=-90,in=90,  looseness=1] (join2);

\end{tikzpicture}
\caption{Cascaded-regression pipeline. The input $\tau(l,d_h,d_v,f)$ splits into horizontal (blue) and vertical (red) components and is finally multiplied to form $\hat{\tau}(l,d_h,d_v,f)$.}
\label{fig:cascaded_regression_pipeline}
\vspace{-5.5mm}
\end{figure}
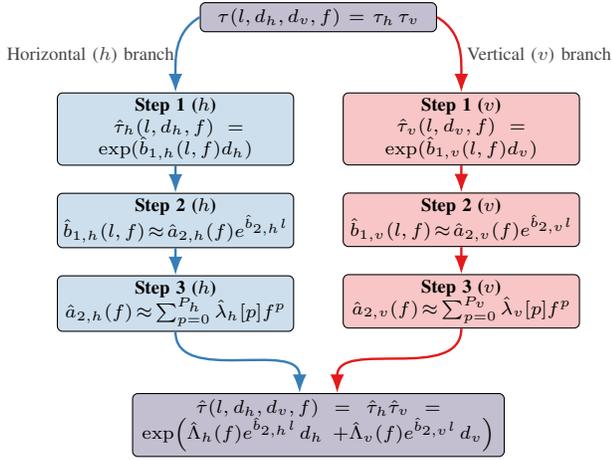

\vspace{1mm}
Finally, the coefficients $\hat{a}_{2,h}(f)$ and $\hat{a}_{2,v}(f)$ obtained in the previous step are approximated by polynomial functions of frequency.
  Let $\Lambda_\xi(f) \triangleq \sum_{i=0}^{P_\xi} \boldsymbol{\lambda}_{\xi}[i] f^{i},$ for $\xi \in \{h, v\}$ we have: 
 \footnotesize
\begin{equation}\label{eq_cascadedReg_3}
\hat{\boldsymbol{\lambda}}_\xi
= \mathop{\mathrm{arg\,min}}_{\boldsymbol{\lambda}_\xi}
\sum_{k}
\left\|
\hat{a}_{2,\xi}\!\left(\boldsymbol{f}[k]\right)
-
\Lambda_{\xi}\!\left(\boldsymbol{f}[k]\right)
\right\|_2^2,
\end{equation}
\normalsize
where ${\boldsymbol{\lambda}}_{\xi}$ denotes the vector of polynomial coefficients for $\xi \in \{h, v\}$, and $\boldsymbol{f}$ is the vector of frequency samples.
Similarly, we define $\hat{\Lambda}_h(f)$ as a degree-$P_{\xi}$ polynomial in $f$ whose coefficients $\hat{\boldsymbol{\lambda}}$ are obtained from~\eqref{eq_cascadedReg_3}. Using this representation, Fig.~\ref{fig:cascaded_regression_pipeline} depicts the cascaded-regression pipeline employed in this paper to model the transmittance $\tau$ as a function of
 $(l,d_h,d_v,f)$.

By solving these three cascaded regressions and substituting the results from \eqref{eq_cascadedReg_3} into \eqref{eq_cascadedReg_2}, and then into \eqref{eq_cascadedReg_1}, we obtain the following expression for the approximated transmittance:

\vspace{-2mm}\footnotesize
\begin{equation}
\label{eq:approx_transmittance}
\hat{\tau}(l,d_h,d_v,f)
 = \exp\!\left(
  \hat{\Lambda}_h(f) e^{\hat{b}_{2,h} l}\, d_h \right.\left. +
  \hat{\Lambda}_v(f) e^{\hat{b}_{2,v} l}\, d_v
\right).
\end{equation}
\normalsize
\section{Data and Implementation Approaches} \label{section_ImpApp}
This section describes the generation of \textit{am} transmittance data for the Dr2Dr, MAAC, and U2U scenarios, and then presents and discusses two alternative implementations of the path-loss estimation model introduced in Section~\ref{section_sysModel}.
\subsection{Atmospheric Model Data}\label{subsection_Data}\vspace{-1mm}
To evaluate the proposed THz path-loss model, reference data is generated using the \textit{am} tool by varying the water-vapor volume mixing ratio, transmitter altitude $z_1=l$, receiver altitude $z_2=l+d\cos\theta$, and zenith 
angle $\theta$ over the discrete grid, under the U.S. Standard Atmosphere 1976 weather profile~\cite{NASA1976}.

For each configuration, the transmittance values calculated by \textit{am} across the different THz and sub-THz frequency bands in \(0.1\text{--}1\,\text{THz}\) are recorded. Next, these outputs are  stored in a four-dimensional matrix denoted by \( \Tau^{am}\), where the $(p,q,r,s)$-th element corresponds to  $\tau^{am}{(\boldsymbol{l}[p],\boldsymbol{d}[q],\boldsymbol{\theta}[r],\boldsymbol{f}[s])}$,
in which the superscript $am$ denotes the tool by which the transmittance values are attained.
\( \boldsymbol{l} \), and 
\( \boldsymbol{d} \)  are the vectors of altitude and distance instances in meters defined differently in each communication scenario, and
\( \boldsymbol{\theta} = [0^\circ, 4.5^\circ, \dots, 90^\circ] \),
respectively. Note that \( \theta = 0^\circ \) and \( \theta = 90^\circ \) represent vertical and horizontal communication scenarios, respectively. $\mathrm{L}^{am}_{\mathrm{total}}$ is defined similarly as the four-dimensional total path loss matrix, according to \eqref{eq_pathLoss_2}.

Moreover, the frequency samples $\boldsymbol{f}[s]$ are taken at every $0.3$ GHz across seven representative sub-THz and three low-THz sub-bands~(Table~\ref{tab:bands}), are selected to exclude sharp molecular absorption peaks primarily caused by $\mathrm{H_2O}$ and $\mathrm{O_2}$ across all altitudes of interest.
Furthermore, to account for different atmospheric conditions and align with practical aerial use cases, three different communication scenarios are considered, namely Dr2Dr, MAAC, and high altitude U2U. The discrete ranges of altitudes and link distances considered for each aerial communication scenario, along with brief descriptions, are summarized in Table~\ref{tab:scenarios}.

\begin{table}[t]
\centering
\vspace{0.5mm}
\caption{Proposed Sub-bands Across 0.1--1 THz}
\label{tab:bands}
\renewcommand{\arraystretch}{1.08}
\setlength{\tabcolsep}{6pt}
\begin{tabular}{clc}
\hline
\textbf{Range} & \textbf{Sub-band Name(s)} & \textbf{Frequency Span [THz]} \\ \hline

\multirow{7}{*}{\shortstack{sub-THz\\($0.100$--$0.750$~THz)}}
 & $\mathrm{D}$--$\mathrm{G}$  & $0.120$--$0.300$ \\
 & $\mathrm{Y0}$      & $0.327$--$0.368$ \\
 & $\mathrm{Y1}$      & $0.386$--$0.423$ \\
 & $\mathrm{Y2}$      & $0.454$--$0.470$ \\
 & $\mathrm{WR0}$     & $0.493$--$0.525$ \\
 & $\mathrm{WR1}$     & $0.594$--$0.618$ \\
 & $\mathrm{WR2}$     & $0.625$--$0.710$ \\ \hline

\multirow{3}{*}{\shortstack{THz\\($\geq 0.750$~THz)}} 
 & $\mathrm{THz0}$    & $0.790$--$0.830$ \\
 & $\mathrm{THz1}$    & $0.836$--$0.910$ \\
 & $\mathrm{THz2}$    & $0.920$--$0.960$ \\ \hline

\end{tabular}
\vspace{-3mm}
\end{table}

\begin{table*}[t]
\centering
\caption{Considered Aerial Communication Scenarios and Parameter Ranges}
\label{tab:scenarios}
\renewcommand{\arraystretch}{1.1}
\setlength{\tabcolsep}{4.5pt}
\begin{tabular}{lccc}
\hline
\textbf{Scenario} & \textbf{Altitude Range ($\boldsymbol{l}$) in km} & \textbf{Distance Range ($\boldsymbol{d}$) in km} & \textbf{Description} \\ \hline
\textbf{Dr2Dr} & 
$[0, 0.01, 0.02, \dots, 0.50]$  &
$[0.01, 0.02, \dots, 0.10]$   &
Low-altitude drone-to-drone communication near ground level. \\[0.6ex]
\textbf{MAAC} & 
$[1, 1.5, 2, \dots, 15]$ &
$[0.5, 1, 1.5, \dots, 10]$ &
Mid-altitude aerial communication between UAVs or aircraft. \\[0.6ex]
\textbf{U2U} & 
$[15,15.5,16,\dots,50]$ &
$[0.5,1,1.5,\dots,50]$ &
High-altitude UAV-to-UAV and HAPS-to-HAPS communication~\cite{saeed2021point}. \\ \hline
\end{tabular}
\vspace{-3mm}
\end{table*}

In order to validate our proposed model using the transmittance data acquired with the \textit{am} tool, we employ two different approaches, namely, $\theta$-agnostic and $\theta$-adaptive, which are elaborated in the sequel.  

\subsection{\texorpdfstring{$\theta$}{theta}-agnostic approach}  
For a target frequency band, the complete \textit{am} transmittance dataset $\Tau^{am}$ within that band is used to compute optimal coefficients via \eqref{eq_cascadedReg_1}–\eqref{eq_cascadedReg_3}. The estimation involves three steps:  
\begin{enumerate}  
    \item \textbf{Step 1:} For each altitude $l$ and frequency, model transmittance exponentially with respect to horizontal ($d_h = d \sin{\theta}$) and vertical ($d_v = d \cos{\theta}$) distances: 
    \vspace{-1mm}
    \begin{equation} \label{NA_eqStep1}  
        \tau(l,d,\theta,f) \approx \hat{\tau}_h \hat{\tau}_v = e^{\hat{b}_{1,h} d_h + \hat{b}_{1,v} d_v}.
    \end{equation}
    \item \textbf{Step 2:} For each frequency, and $\xi\in\{h,v\}$, fit $\hat{b}_{1,\xi}(l,f)$, exponentially with respect to altitude: $ \hat{b}_{1,\xi}(l,f) \approx \hat{a}_{2,\xi} e^{\hat{b}_{2,\xi} l}$,  
    where $\hat{b}_{2,\xi}(f)$ is nearly constant and replaced by its mean over the band.  
    \item \textbf{Step 3:} Fit $\hat{a}_{2,\xi}(f)$ with degree-$P_\xi$ polynomial $\hat{\Lambda}_\xi(f)$.
\end{enumerate}  

\subsection{\texorpdfstring{$\theta$}{theta}-adaptive approach}
Here, the cascaded regression is independently fitted for each zenith angle $\theta$ over its discrete range. For each $\theta$, the dataset $\tau^{am}(l,d,\theta,f)$ is used without separating horizontal and vertical components, since $d_h$ and $d_v$ are proportional for a fixed $\theta$. Define $\tau_{\theta}(l,d,f)$ as the transmittance values when the zenith angle is fixed to $\theta$. Then, the cascaded regression steps \eqref{eq_cascadedReg_1}--\eqref{eq_cascadedReg_3} are as follows:  
\begin{enumerate}  
    \item \textbf{Step 1:} For each altitude $l$ and frequency $f$, model $\tau_{\theta}(l,d,f)$ exponentially with respect to $d$:  
    \begin{equation} \label{TA_eqStep1}  
        \tau_{\theta}(l,d,f) \approx e^{(\hat{b}_{1,\theta,h}\sin{\theta}+\hat{b}_{1,\theta,v}\cos{\theta})d} = e^{\hat{b}_{1,\theta}d}.  
    \end{equation}    
    \item \textbf{Step 2:} Fit $\hat{b}_{1,\theta}(l,f)$ exponentially with respect to altitude:  $        \hat{b}_{1,\theta}(l,f) \approx \hat{a}_{2,\theta} e^{\hat{b}_{2,\theta} l}$ with $\hat{b}_{2,\theta}(f)$ set to its mean over $f$.  
    \item \textbf{Step 3:} Fit $\hat{a}_{2,\theta}(f)$ with a polynomial of degree $P_\theta$: $\hat{a}_{2,\theta}(f) \approx  \hat{\Lambda}_\theta(f)$
\end{enumerate}  
This process is repeated for all $\theta$ samples. Assuming $P_\theta = P$ for all angles, the total number of coefficients is $N_\theta(P+2)$, where $N_\theta$ is the number of zenith-angle samples.

\subsection{Complexity Comparison} 
In terms of computational complexity, the $\theta$-agnostic 
approach performs the cascaded regression once over the entire 
dataset, whereas $\theta$-adaptive estimates coefficients 
separately for each $\theta$ (i.e., $N_\theta$ times), each using 
$1/N_\theta$ of the samples. Due to the reduction in~\eqref{TA_eqStep1}, the 
$\theta$-adaptive estimator admits a simpler expression than 
that of $\theta$-agnostic approach. 
Let $N_s$, $N_l$, $N_d$, $N_\theta$, and $N_f$ denote the 
numbers of transmittance, altitude, distance, zenith-angle, 
and frequency samples, respectively; the computational 
orders are
\begin{equation} \label{eq_CompO}
\begin{aligned}
\mathcal{C}_{\mathrm{agn}} &= O\big(N_s + N_l N_f + N_f(P_h^2+P_v^2)\big),\\
\mathcal{C}_{\mathrm{adap}} &= O\big(N_s + N_\theta N_l N_f + N_\theta N_f P^2\big),
\end{aligned}
\end{equation}
where $N_s = N_l N_d N_\theta N_f$.
The three additive terms in~\eqref{eq_CompO} correspond, respectively, to the first, second, and third regression steps in each corresponding approach. Both methods, therefore, scale as $O(N_s)$ for large datasets, although the $\theta$-adaptive approach has larger multiplicative constants (especially for large $N_\theta$) due to separate regressions per zenith-angle sample. A comparative summary of the two approaches, highlighting coefficient scaling, asymptotic complexity, and closed-form dependence on the zenith angle, is given in Table~\ref{tab:comparison}.
\begin{table}[!ht]
\vspace{-2mm}
\caption{Complexity of $\theta$-agnostic and $\theta$-adaptive approaches}
\centering
\setlength{\tabcolsep}{3.5pt}
\begin{tabular}{|c|c|c|c|}
\hline
\textbf{Approach} 
& \textbf{\# Coefficients} 
& \textbf{Complexity Order} 
& \textbf{Closed-form in $\theta$} \\
\hline
$\theta$-agnostic 
& $P_h + P_v + 4$ 
& $O(N_s)$ 
& Yes \\
\hline
$\theta$-adaptive 
& $N_\theta (P + 2)$ 
& $O(N_s)$ 
& No \\
\hline
\end{tabular}
\label{tab:comparison}
\vspace{-1mm}
\end{table}
\vspace{-2mm}
\section{Numerical Results and Discussion} \label{section_Results}
The two proposed THz path loss estimation methods are optimized using the \textit{am}-generated transmittance datasets for the considered aerial scenarios. Their performance is evaluated and compared using the Normalized Root Mean Square Error (NRMSE):
Letting $\hat{\mathrm{L}}^{am}_{\mathrm{total}}$ be the approximated total path loss matrix using an estimation method, the root mean square error (RMSE) in dB and NRMSE for that estimation method are defined as

\vspace{-1mm}
\footnotesize
\begin{equation}
\begin{split}\label{eq_RMSE_NRMSE}
 \mathrm{RMSE}  &= \frac{1}{\sqrt{N_{s}}}\lVert \hat{\mathrm{L}}^{am}_{\mathrm{total}}- \mathrm{L}^{am}_{\mathrm{total}} \rVert_{F},\\
 \mathrm{NRMSE} &= \frac{\mathrm{RMSE\,}}{\frac{1}{N_s}\sum_{p,q,r,s}{L_{\mathrm{total}}(\boldsymbol{l}[p],\boldsymbol{d}[q],\boldsymbol{\theta}[r],\boldsymbol{f}[s])}}, 
\end{split} 
\end{equation}
\normalsize
 where, $\lVert \mathrm{X}\rVert_{F}$ is the Frobenius norm of matrix $\mathrm{X}$. It is important to note that, in all results, the polynomial degrees are fixed to 6 for both the vertical and horizontal components in the \Tag~approach and for each zenith-angle instance in the \Tad~approach. This specific choice is based on preliminary experiments showing that it provides generally accurate and consistent results across all considered cases.
\begin{figure*}[t]
\centering
\subfloat[D--G sub-band]{\includegraphics[width=0.24\textwidth]{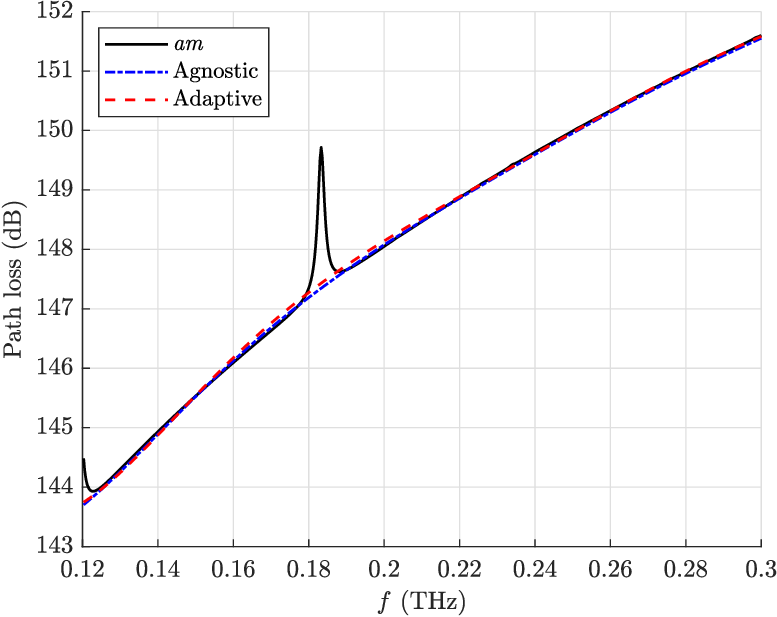}}
\hfill
\subfloat[Y-band]{\includegraphics[width=0.24\textwidth]{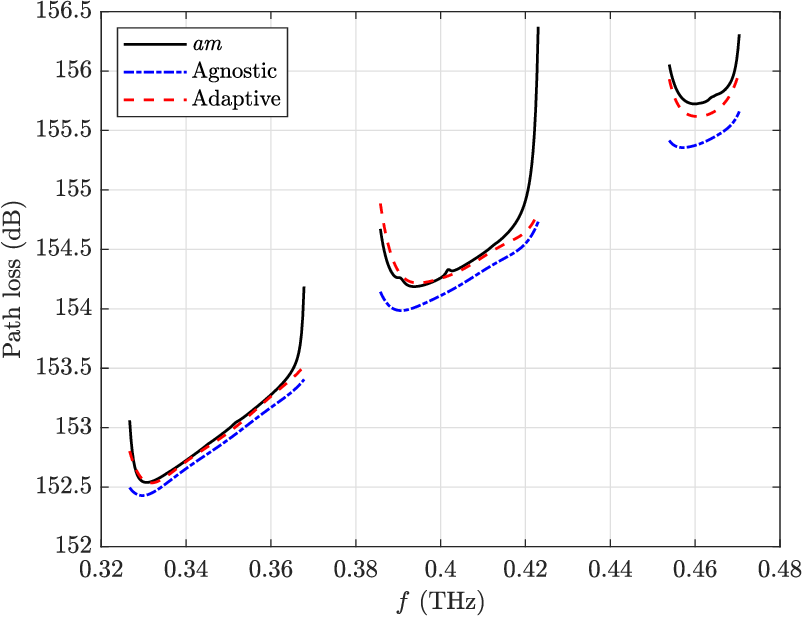}}
\hfill
\subfloat[WR-1.5 band]{\includegraphics[width=0.24\textwidth]{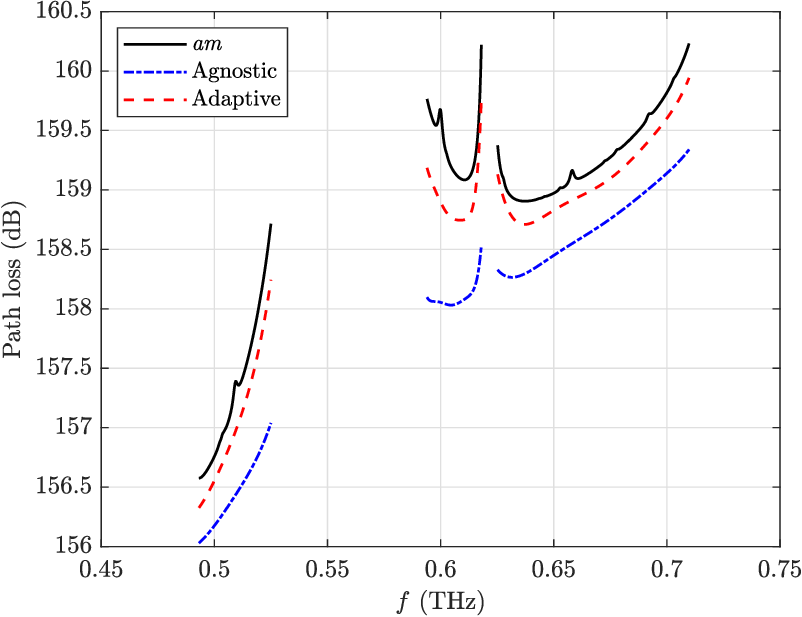}}
\hfill
\subfloat[THz band]{\includegraphics[width=0.24\textwidth]{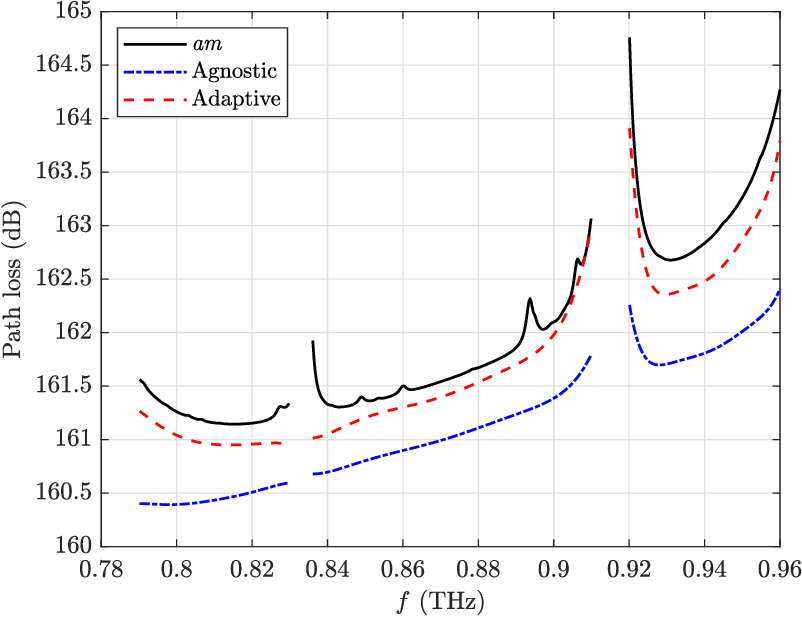}}
\caption{Example path loss curves from the \textit{am}-tool and the proposed $\theta$-agnostic and $\theta$-adaptive estimates for $l=8$ km, $d=3$ km, and $\theta=9^{\circ}$ in the MAAC scenario, shown across all sub-bands.}
\label{fig:MAAC_pathloss}
\vspace{-4mm}
\end{figure*}

\begin{figure*}[t]
 \centering
 \subfloat[NRMSE vs $d$ \label{subfig:Dr2Dr_b}]{\includegraphics[width=0.32\textwidth]{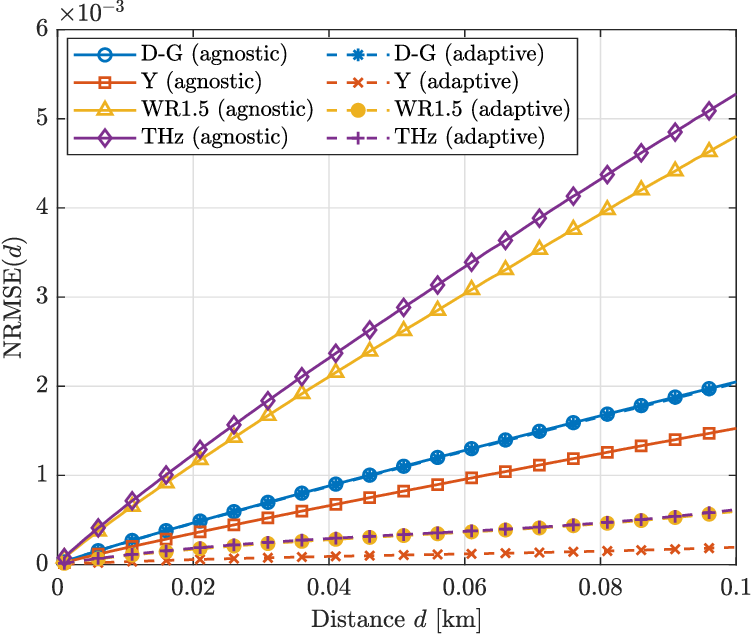}}
 \hfill
 \subfloat[NRMSE vs $l$ \label{subfig:Dr2Dr_c}]{\includegraphics[width=0.327\textwidth]{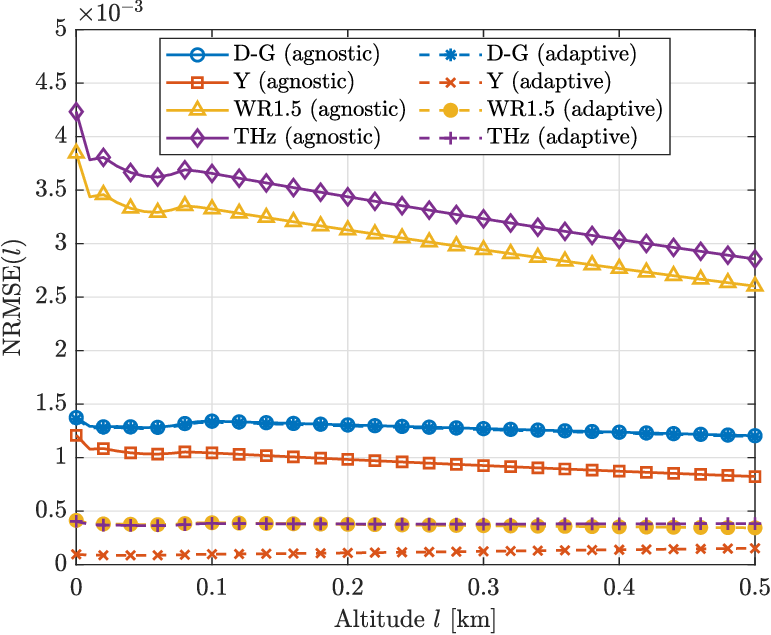}}
 \hfill
 \subfloat[NRMSE vs $\theta$ \label{subfig:Dr2Dr_d}]{\includegraphics[width=0.32\textwidth]{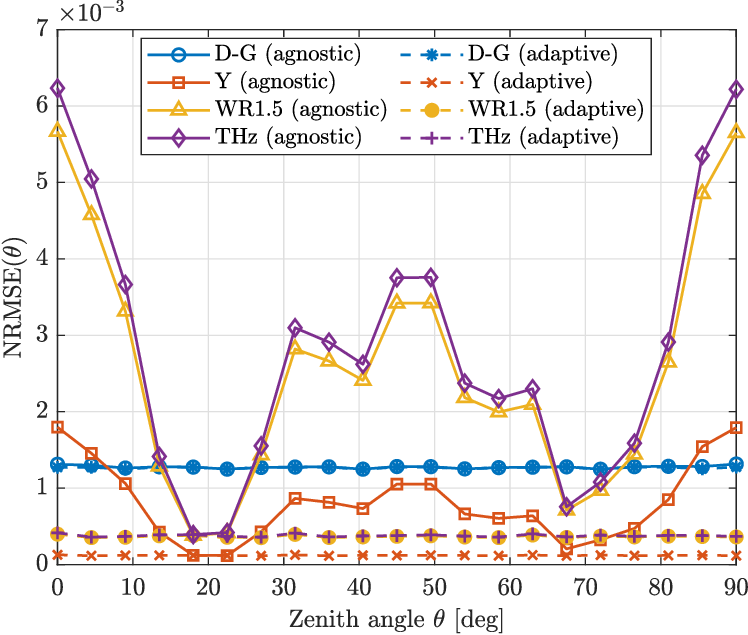}}
\caption{Dr2Dr scenario. NRMSE of the $\theta$-agnostic and $\theta$-adaptive path loss estimations as a function of (a) communication distance $d$ in km; (b) the minimum altitude of the transceivers $l$ in km; and (c) zenith angle $\theta$ in degrees.}
\label{fig:NRMSEcurves_Dr2Dr}
\vspace{-4mm}
\end{figure*}

\begin{figure*}[t]
\centering
\subfloat[NRMSE vs $d$ \label{subfig:MAAC_b}]{\includegraphics[width=0.32\textwidth]{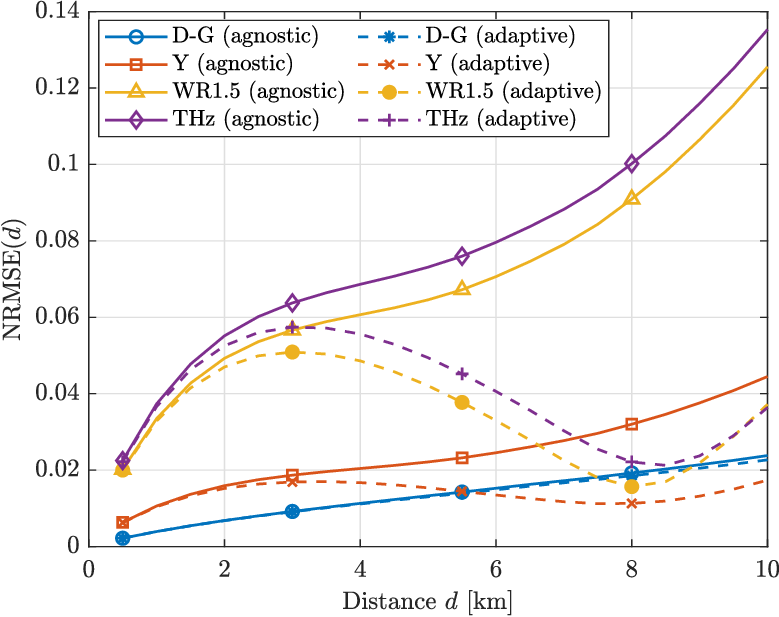}}
\hfill
\subfloat[NRMSE vs $l$ \label{subfig:MAAC_c}]{\includegraphics[width=0.32\textwidth]{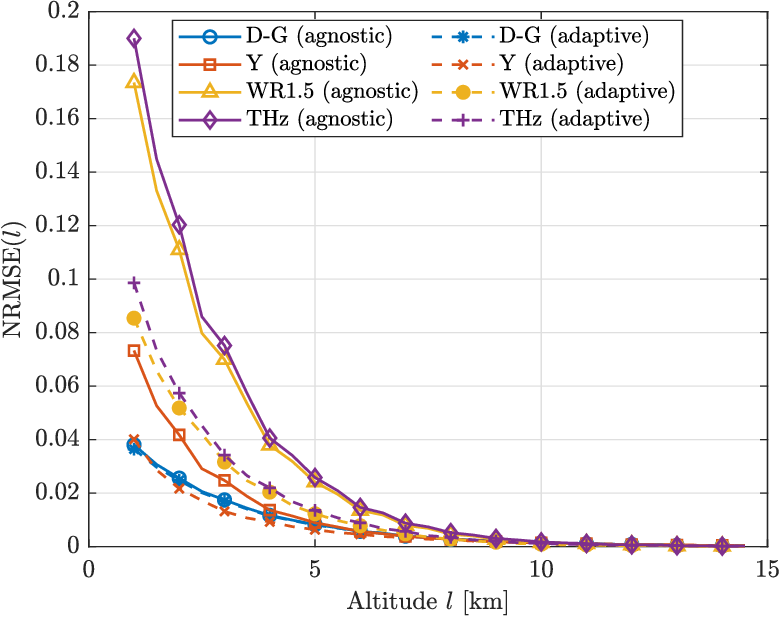}}
\hfill
\subfloat[NRMSE vs $\theta$\label{subfig:MAAC_d}]{\includegraphics[width=0.32\textwidth]{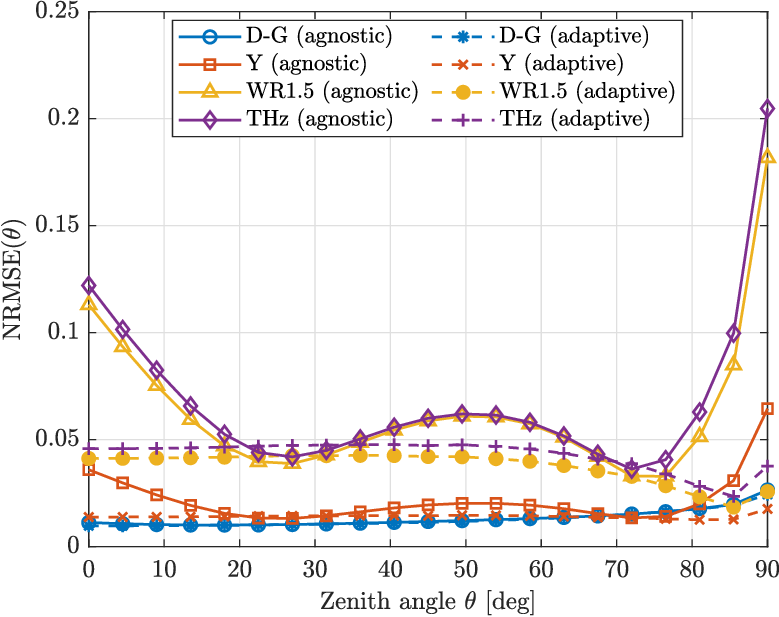}}
\caption{MAAC scenario: NRMSE of the $\theta$-agnostic and $\theta$-adaptive path loss estimations as a function of (a) communication distance $d$ in km; (b) the minimum altitude of the transceivers $l$ in km; and (c) zenith angle $\theta$ in degrees.}
\label{fig:NRMSEcurves_MAAC}
\vspace{-3mm}
\end{figure*}

\begin{figure*}[t]
\vspace{-4mm}
\centering
\subfloat[NRMSE vs $d$\label{subfig:UAV_b}]{\includegraphics[width=0.33\textwidth]{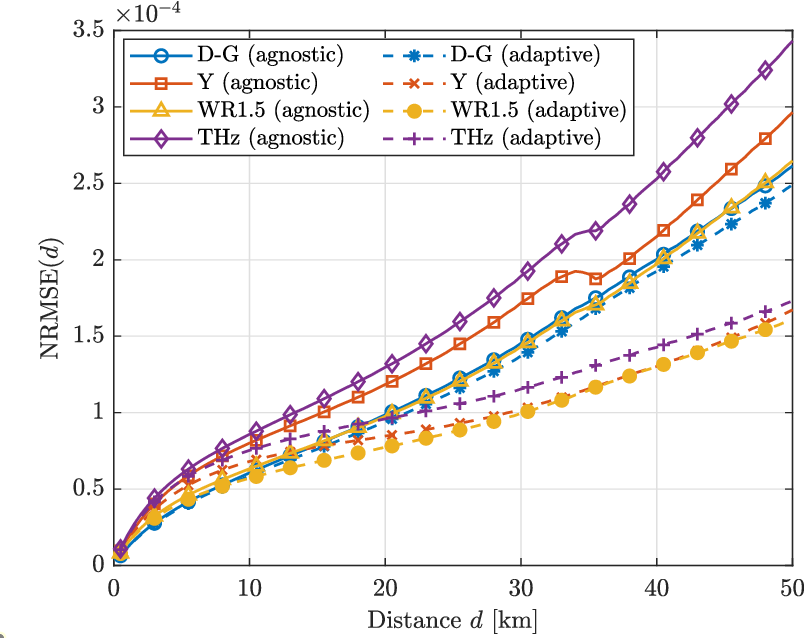}}
\hfill
\subfloat[NRMSE vs $l$\label{subfig:UAV_c}]{\includegraphics[width=0.31\textwidth]{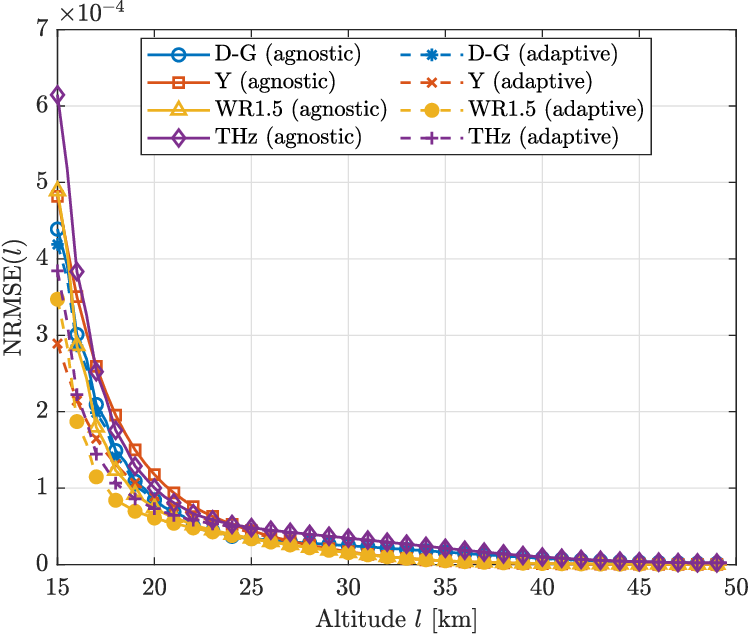}}
\hfill
\subfloat[NRMSE vs $\theta$\label{subfig:UAV_d}]{\includegraphics[width=0.32\textwidth]{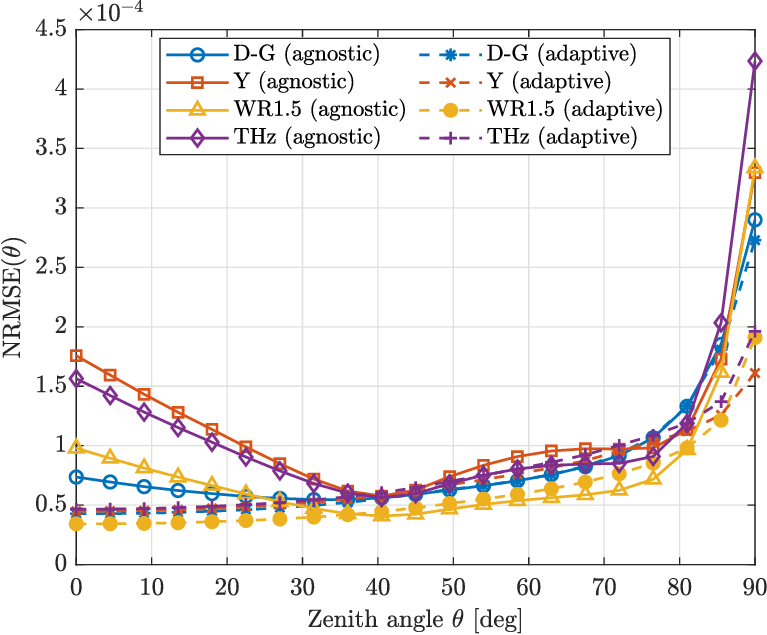}}
\caption{U2U scenario: NRMSE of the $\theta$-agnostic and $\theta$-adaptive path loss estimations as a function of (a) communication distance $d$ in km; (b) the minimum altitude of the transceivers $l$ in km; and (c) zenith angle $\theta$ in degrees.}
\label{fig:NRMSEcurves_UAV}
\vspace{-4mm}
\end{figure*}

Fig.~\ref{fig:MAAC_pathloss} gives a typical example of the path loss curves versus frequency for $l=8$ km, $d=3$ km, and $\theta=9^\circ$ in the MAAC scenario. Both $\theta$-adaptive and $\theta$-agnostic methods closely follow the \textit{am} reference curve, with $\theta$-adaptive being slightly more accurate in this particular example case.
Fig.~\ref{fig:NRMSEcurves_Dr2Dr}--\ref{fig:NRMSEcurves_UAV} illustrate the modeling accuracy over 0.1--1~THz 
for Dr2Dr, MAAC, and U2U scenarios. To analyze estimation 
error, one parameter (altitude, distance, or zenith angle) 
is fixed while NRMSE is computed over the remaining 
dimensions, yielding $\mathrm{NRMSE}(l)$, $\mathrm{NRMSE}(d)$, and $\mathrm{NRMSE}(\theta)$.
For tractability, sub-bands within the Y, WR-1.5, and THz ranges are aggregated.

As shown in Fig.~\ref{fig:NRMSEcurves_Dr2Dr}a--c, NRMSE increases almost linearly with distance, 
with $\theta$-agnostic having a steeper slope than $\theta$-adaptive. 
It decreases with increasing altitude, particularly for 
$\theta$-agnostic, implying a reduced performance gap between the two models at higher altitudes. 
The $\theta$-agnostic model degrades at extreme zenith angles, 
where many polynomial coefficients contribute negligibly, whereas the $\theta$-adaptive approach maintains 
uniformly low and stable $\mathrm{NRMSE}(\theta)$ across all angles.
Fig.~\ref{fig:NRMSEcurves_MAAC} exhibits behavior similar to Dr2Dr, except that 
$\mathrm{NRMSE}(d)$ (Fig.~\ref{subfig:MAAC_b}) is not strictly increasing with distance 
for $\theta$-adaptive in the WR1.5 and THz bands. 
Estimation accuracy decreases in MAAC compared to Dr2Dr, 
mainly due to its wider distance range, which introduces 
greater path loss variability. 
In high-altitude U2U links, the thin atmosphere renders 
absorption nearly negligible relative to FSPL. 
Accordingly, the curves in Fig.~\ref{fig:NRMSEcurves_UAV} converge, with 
$\theta$-adaptive only slightly outperforming 
$\theta$-agnostic. Errors remain low in this regime, making total path loss 
easier to approximate; hence, the less complex 
$\theta$-agnostic approach may be preferable due to its 
marginal performance gap.

\begin{figure*}[t]
 \centering
 \subfloat[Dr2Dr \label{subfig:Dr2Dr_a}]{\includegraphics[width=0.31\textwidth]{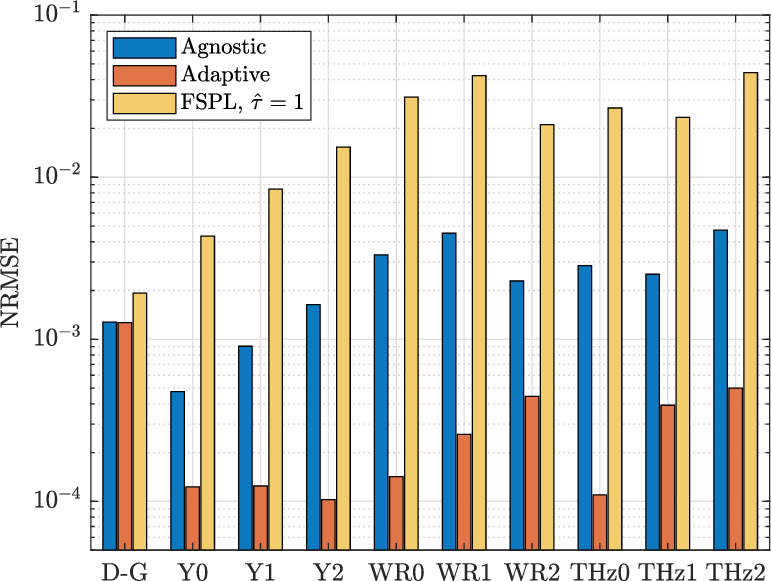}}
 \hfill
 \subfloat[MAAC \label{subfig:MAAC_a}]{\includegraphics[width=0.31\textwidth]{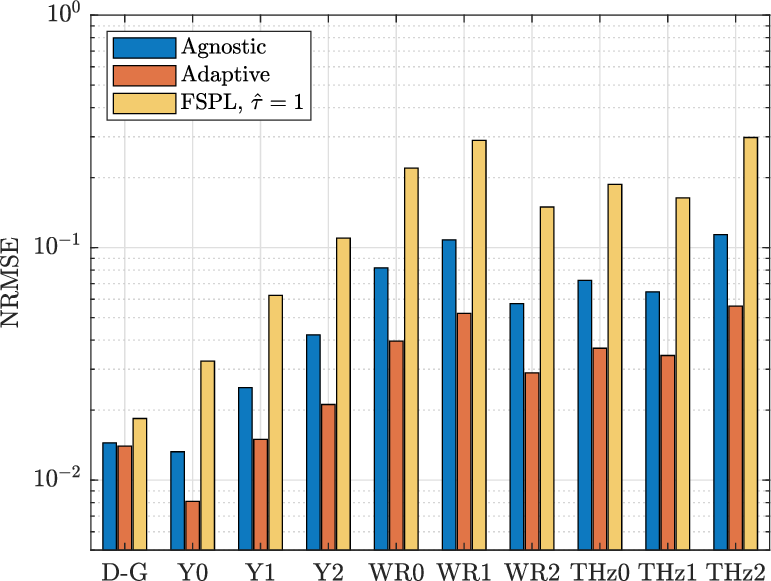}}
 \hfill
 \subfloat[U2U \label{subfig:UAV_a}]{\includegraphics[width=0.31\textwidth]{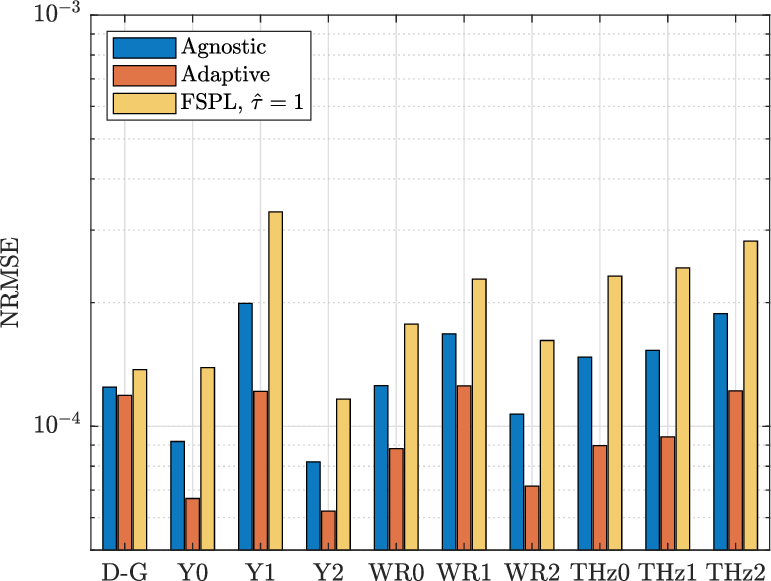}}
\caption{NRMSE of the $\theta$-agnostic, $\theta$-adaptive, and FSPL methods across all considered sub-bands in (a) Dr2Dr, (b) MAAC, and (c) U2U aerial communication scenarios.}
\label{fig:NRMSEbarplots}
\vspace{-3mm}
\end{figure*}

Fig.~\ref{fig:NRMSEbarplots} shows the NRMSE of the proposed methods across all sub-bands for each communication scenario. As a benchmark, the FSPL-only model (assuming no absorption) is also included for comparison.
In the Dr2Dr scenario (Fig.~\ref{subfig:Dr2Dr_a}), \Tad~outperforms \Tag, achieving nearly an order of magnitude lower NRMSE across all sub-bands. \Tag~also improves over the FSPL baseline, except in the D--G band where absorption is minimal aside from a narrow peak near $0.183$~THz.
In the MAAC scenario (Fig.~\ref{subfig:MAAC_a}), \Tad~still performs better, though the gap shrinks due to reduced absorption at mid altitudes, bringing both methods closer to the FSPL baseline.
Finally in U2U scenario (Fig.~\ref{subfig:UAV_a}), all errors converge, with \Tad~slightly outperforming \Tag, and \Tag~closely matching the FSPL model, reaffirming that the more complex \Tad~approach may be unnecessary in this case.
\section{Conclusions}\label{section_Conclusion}
This paper has presented a new analytical path loss model for THz-band aerial communications, capable of capturing arbitrary 3D transceiver geometries and frequency-selective absorption. The model is validated using reliable atmospheric transmittance data generated by the \textit{am} tool for multiple aerial scenarios, including Dr2Dr, MAAC, and U2U links across ten different representative bands in the 0.1–1 THz range. Two implementation approaches, namely \Tag~and \Tad~have been developed and compared. The results have demonstrated that the \Tad~approach achieves higher accuracy, especially in low-altitude scenarios, \textit{e.g.,} Dr2Dr, where absorption is dominant. On the other hand, the \Tag~approach has offered comparable performance with reduced complexity at higher altitudes \textit{e.g.,} high altitude U2U links.
Overall, the proposed framework provides a unified and accurate basis for THz path loss analysis and can guide the design of future high-frequency aerial communication systems.

From a practical perspective, the proposed model supports link budget analysis and coverage planning by providing a closed-form expression of path loss as a function of transceiver locations, without requiring repeated numerical simulations. This framework can be particularly useful for space–air-ground integrated networks, mobility-aware path tracking, and adaptive beamforming, where geometry varies dynamically and fast path loss updates are required. Future work may extend the model to account for varying weather conditions and to validate or retrain the regression parameters using experimental propagation measurements when available.

\bibliographystyle{IEEEtran}
\bibliography{references}

@article{AnalyticalModel_paper,
  title={A simple analytical model for {THz} band path loss},
  author={Erdem, Mikail and Saleem, Ammar and Gurbuz, Ozgur and Ecemis, Cansu and Saeed, Akhtar},
  journal={IEEE Communications Letters},
  volume={27},
  number={3},
  pages={996--1000},
  year={2023},
  publisher={IEEE}
}

@misc{am_Tool,
  author       = {Paine, Scott},
  title        = {The am atmospheric model},
  month        = jul,
  year         = 2022,
  publisher    = {Zenodo},
  version      = {12.2},
  doi          = {10.5281/zenodo.6774376},
  url          = {https://doi.org/10.5281/zenodo.6774376},
}

@techreport{NASA1976,
  author       = {NASA},
  title        = {U.{S}. Standard Atmosphere, 1976},
  institution  = {National Aeronautics and Space Administration (NASA)},
  address      = {Washington, D.C.},
  year         = {1976},
  number       = {NASA-TM-X-74335},
  note         = {{U}.S. Government Printing Office},
  url          = {https://ntrs.nasa.gov/citations/19770009539}
}

@ARTICLE{6GZhang,
  author={Zhang, Zhengquan and Xiao, Yue and Ma, Zheng and Xiao, Ming and Ding, Zhiguo and Lei, Xianfu and Karagiannidis, George K. and Fan, Pingzhi},
  journal={IEEE Vehicular Technology Magazine}, 
  title={{6G} Wireless Networks: Vision, Requirements, Architecture, and Key Technologies}, 
  year={2019},
  volume={14},
  number={3},
  pages={28-41},
  doi={10.1109/MVT.2019.2921208}}

@article{taleb2023propagation,
  title={{Propagation of {T}{H}z radiation in air over a broad range of atmospheric temperature and humidity conditions}},
  author={Taleb, Fatima and Alfaro-Gomez, Mariana and Al-Dabbagh, Mohanad Dawood and Ornik, Jan and Viana, Juan and J{\"a}ckel, Alexander and Mach, Cornelius and Helminiak, Jan and Kleine-Ostman, Thomas and K{\"u}rner, Thomas and others},
  journal={Scientific Reports},
  volume={13},
  number={1},
  pages={20782},
  year={2023},
  publisher={Nature Publishing Group UK London}
}

@article{liao2023attenuation,
  title={Attenuation characterization of terahertz waves in foggy and rainy Conditions at 0.1--1 {THz} Frequencies},
  author={Liao, Xi and Fan, Linjie and Wang, Yang and Yu, Ziming and Wang, Guangjian and Li, Xianjin and Zhang, Jie},
  journal={Electronics},
  volume={12},
  number={7},
  pages={1684},
  year={2023},
  publisher={MDPI}
}

@article{SAEED_Phycom,
author = {Saeed, Akhtar and Gurbuz, Ozgur and Akkaş, Mustafa},
year = {2020},
month = {05},
pages = {101113},
title = {Terahertz communications at various atmospheric altitudes},
volume = {41},
journal = {Physical Communication},
doi = {10.1016/j.phycom.2020.101113}
}

@ARTICLE{TMao2022,
  author={Mao, Tianqi and Zhang, Leyi and Xiao, Zhenyu and Han, Zhu and Xia, Xiang-Gen},
  journal={IEEE Communications Magazine}, 
  title={Terahertz-Band Near-Space Communications: From a Physical-Layer Perspective}, 
  year={2024},
  volume={62},
  number={2},
  pages={110-116},
  doi={10.1109/MCOM.004.2200429}}

@article{jornet2011channel,
  title={Channel modeling and capacity analysis for electromagnetic wireless nanonetworks in the {T}erahertz band},
  author={Jornet, Josep Miquel and Akyildiz, Ian F},
  journal={IEEE Transactions on Wireless Communications},
  volume={10},
  number={10},
  pages={3211--3221},
  year={2011},
  publisher={IEEE}
}

@article{saeed2021point,
  title={Point-to-point communication in integrated satellite-aerial 6{G} networks: State-of-the-art and future challenges},
  author={Saeed, Nasir and Almorad, Heba and Dahrouj, Hayssam and Al-Naffouri, Tareq Y and Shamma, Jeff S and Alouini, Mohamed-Slim},
  journal={IEEE Open Journal of the Communications Society},
  volume={2},
  pages={1505--1525},
  year={2021},
  publisher={IEEE}
}

@article{kokkoniemi2018simplified,
  title={Simplified molecular absorption loss model for 275–400 {GHz} frequency band},
  author={Kokkoniemi, Jukka and Petrov, Viktor and Moltchanov, Dmitri and K\"{a}rkk\"{a}inen, Mika and Melvasalo, Joonas and Koucheryavy, Yevgeni and J\"{a}ntti, Riku and Latva-aho, Matti},
  journal={IEEE Access},
  volume={6},
  pages={5700--5710},
  year={2018},
  doi={10.1109/ACCESS.2018.2794484}
}

@article{danobrega2023channel,
  title={A Channel Loss Model for {T}{H}z Networks From 100–600 {GHz} Considering Both Molecular and Water Vapor Continuum Absorptions},
  author={da N{\'o}brega, Ricardo and Silva, R. and Vieira, J. and Maciel, T.},
  journal={IEEE Transactions on Communications},
  volume={71},
  number={9},
  pages={5631--5645},
  year={2023},
  doi={10.1109/TCOMM.2023.3271512}
}

@inproceedings{khan2025analysis,
  title={Analysis of Clouds and Rain Losses on {T}erahertz Band for Non-Terrestrial Networks},
  author={Khan, Bushra and Kokkoniemi, Joonas},
  booktitle={2025 19th European Conference on Antennas and Propagation (EuCAP)},
  pages={1--5},
  year={2025},
  organization={IEEE}
}

@article{doborshchuk2022propagation,
  title={Propagation model for ground-to-aircraft communications in the {T}erahertz band with cloud impairments},
  author={Doborshchuk, Vladimir and Begishev, Vyacheslav and Samouylov, Konstantin},
  journal={Energies},
  volume={15},
  number={21},
  pages={8022},
  year={2022},
  publisher={MDPI}
}

@inproceedings{saeed2024terahertz,
  title={Terahertz drone communications under fog and beam misalignment},
  author={Saeed, Akhtar and Gurbuz, Ozgur},
  booktitle={2024 7th International Balkan Conference on Communications and Networking (BalkanCom)},
  pages={43--47},
  year={2024},
  organization={IEEE}
}

\end{document}